\documentclass[a4paper,12pt]{article}
\usepackage{epsfig}
\usepackage{subfigure}
\usepackage{amssymb, graphics}
\usepackage{pslatex}
\usepackage{graphicx}
\usepackage{bbm}
\usepackage{latexsym}
\usepackage{amsfonts}
\usepackage{amsmath}
\usepackage[title,titletoc]{appendix}
\usepackage[super]{nth}
\usepackage{cancel}

\newcommand{\mathsym}[1]{{}}

\newcommand{\qed}{\nobreak \ifvmode \relax \else \ifdim\lastskip<1.5em \hskip-\lastskip \hskip1.5em plus0em minus0.5em 
\fi \nobreak \vrule height0.75em width0.5em depth0.25em\fi}

\def\app#1#2{  \mathrel{    \setbox0=\hbox{$#1\sim$}    \setbox2=\hbox{      \rlap{\hbox{$#1\propto$}}     
 \lower1.1\ht0\box0    }    \raise0.25\ht2\box2  }}

\begin{document}

\begin{titlepage}
\begin{center}
\hfill    CERN-PH-TH-2015-012\\

\vskip 1cm

{\large \bf Natural Quasi-Alignment with two Higgs Doublets and RGE Stability}

\vskip 1cm

F. J. Botella$^{a,}$ \footnote{E-mail: fbotella@uv.es},
G. C. Branco$^{b,c,d,}$\footnote{E-mail: gbranco@tecnico.ulisboa.pt}, 
Ant\'onio M. Coutinho$^{b,}$\footnote{E-mail: 
antonio.coutinho@tecnico.ulisboa.pt},
M. N. Rebelo$^{b,}$\footnote{E-mail: rebelo@tecnico.ulisboa.pt} and
J. I. Silva-Marcos$^{b,}$\footnote{E-mail: juca@cftp.ist.utl.pt}

\vskip 0.07in

{\em 
$a$ Departament de F{\'\i}sica Te\`orica and IFIC, \\
Universitat de Val\`encia - CSIC, E-46100, Burjassot, Spain \\
\vskip 0.5cm

$b$ Centro de F{\'\i}sica Te\'orica de Part{\'\i}culas - CFTP, \\
$c$ Departamento de F\'{\i}sica,\\}
\vskip 0.5cm

{\it  Instituto Superior T\'ecnico - IST, Universidade de Lisboa - UL, }
\\
{\it
Avenida Rovisco Pais, 1049-001 Lisboa, Portugal}
\vskip 0.5cm

{\em 
$d$ Theory Group, Physics Department, CERN, CH-1211, Geneva 23, Switzerland }
\end{center}

\begin{abstract}
In the context of two Higgs doublet models, we study the conditions
required in order to have stable quasi-alignment in flavour space. We show that stability 
under the RGE imposes strong constraints on the flavour structure of the Yukawa 
couplings associated to each one of the Higgs doublets. In particular, we find a novel 
solution, where all Yukawa couplings are proportional to the so-called democratic matrix. This solution is rather unique, since it is the only stable solution which 
is a good starting point for reproducing the observed pattern of quark masses and
mixing. We also show that this stable solution can be obtained by imposing on
the Lagrangian a $Z_3 \times Z^\prime_3$ flavour symmetry. Quark masses of 
the lighter quark generations are generated through the breaking of this
discrete symmetry, and, at this stage, scalar mediated FCNC arise, but are
naturally suppressed by the smallness of light quark masses. In this way,
we relate Higgs alignment to the hierarchy of quark masses
through a discrete family symmetry.
\end{abstract}

\end{titlepage}

\newpage

\section{Introduction}

One of the simplest extensions of the Standard Model (SM) consists of the
addition of scalar doublets to the SM spectrum. Multi-Higgs extensions arise
in a variety of frameworks, including supersymmetric extensions of the SM,
as well as models with family symmetries. A two Higgs doublet model (2HDM) was
first introduced by Lee \cite{Lee:1973iz}, in order to achieve spontaneous
breaking of the CP symmetry. If no extra symmetries are introduced, 2HDMs
lead to too large tree level scalar mediated
flavour-changing-neutral-currents (FCNC) \cite{Branco:2011iw}. 
In order to avoid these
potentially dangerous currents, various schemes have been proposed: \newline

i) Glashow and Weinberg \cite{Glashow:1976nt} have pointed out that one can
avoid FCNC at tree level by introducing a $Z_2$ symmetry under which the two
Higgs doublets transform differently. The introduction of a $Z_2$
symmetry in 2HDMs prevents the generation of spontaneous 
CP breaking \cite{Branco:1980sz}
unless the symmetry is softly broken \cite{Branco:1985aq}.  \\
ii) Pich and Tuzon \cite{Pich:2009sp} have conjectured the existence of flavour
alignment of the two Yukawa matrices, thus avoiding FCNC at tree level. This
is an interesting suggestion, but it has the drawback of being an ad-hoc
assumption, not explained by any symmetry.  Furthermore, it has been pointed out
that in general this scheme is not stable under the 
renormalization group \cite{Ferreira:2010xe}. There have been attempts at obtaining 
alignment in various extensions of the SM \cite{Serodio:2011hg}, \cite{Varzielas:2011jr},
\cite{Celis:2014zaa}.\\
iii) Another possibility has been proposed some time ago \cite{Branco:1996bq}
by Branco, Grimus and Lavoura (BGL) who have pointed out that there is a
symmetry which, when imposed on the Lagrangian, constrains the Yukawa
couplings in such a way that FCNC do arise at tree level, but are entirely
determined by the $V_{CKM}$ matrix, with no other free parameters. 
In some of the
BGL models, one has a strong natural suppression of the most dangerous FCNC,
with, for example, the  strangeness changing neutral currents,
proportional to $(V_{td} V_{ts} ^*)^2$, which implies a very strong natural
suppression of the contribution to the $K^0 - \bar{K^0}$ transition. With this
suppression, the neutral Higgs masses need not be too large. BGL models have
been extended to the leptonic sector \cite{Botella:2011ne}, their relation
to Minimal Flavour Violation models has been studied \cite{Botella:2009pq}
and their phenomenological implications have been recently analysed 
\cite{Botella:2014ska}. \newline

In this paper, we reexamine the question of the stability of flavour alignment 
under the renormalization group. Assuming that the Yukawa couplings of the
two Higgs doublets are aligned, i.e., proportional to each other, we study under
what conditions the alignment is maintained by the renormalization group.
Apart from the conditions already found in Ref.\cite{Ferreira:2010xe}, we find new 
solutions  which can be of great physical interest. One of these solutions, 
corresponds to having all the Yukawa coupling matrices proportional to the 
so-called  democratic matrix \cite{Branco:1990fj}, \cite{Fritzsch:1994yj}, 
\cite{Branco:1995pw}.
This solution is rather unique, since on the one 
hand it is stable under the renormalization group equations
(RGE) and on the other hand, it is the only stable solution
which provides a good starting point for reproducing the observed pattern of quark 
masses and mixing. 
We then point out that this flavour democratic solution 
can be obtained as a result of a $Z_3 \times Z^\prime_3$ flavour symmetry.
In the framework that we propose, flavour alignment is exact in the limit 
where only the third family acquires mass. Once the two light generations acquire
a mass, there are small deviations from alignment, which are suppressed by
the strong hierarchy of quark masses. As a result, one
obtains in this framework, a quasi-alignment of the Yukawa couplings, as a
result of the $Z_3 \times Z^\prime_3$ symmetry, together with the strong
hierarchy of quark masses. \newline

The paper is organised as follows. In the next section, we briefly describe
the general flavour structure of the 2HDM, in order to settle our notation.
In section 3 we derive all the solutions for the Yukawa couplings, leading 
to alignment, stable under the renormalization group. In section 4 we show 
that the flavour democratic solution can be obtained as a result of a 
$Z_3 \times Z^\prime_3$ flavour symmetry and  propose an 
ansatz for  the breaking of the $Z_3 \times Z^\prime_3$ symmetry. 
In section 5, we examine the suppression of scalar mediated FCNC
in our framework. In section 6, we perform a numerical analysis,
showing how the pattern of quark masses and mixing can be obtained in
the framework of our ansatz. Finally our conclusions are contained 
in section 7. In the Appendix we present a full study of the solutions of 
the alignment conditions.

\section{Yukawa Couplings in the General Two-Higgs-Doublet-Model (2HDM)}

For completeness and in order to establish our notation we
briefly review the flavour structure of the 2HDM, when no extra symmetries
are introduced in the Lagrangian. The Yukawa couplings can be written: 
\begin{equation}
\mathcal{L}_{Y}=-\overline{Q_{L}^{0}}\ \Gamma _{1}\Phi _{1}d_{R}^{0}-
\overline{Q_{L}^{0}}\ \Gamma _{2}\Phi _{2}d_{R}^{0}-\overline{Q_{L}^{0}}\
\Omega _{1}\tilde{\Phi}_{1}u_{R}^{0}-\overline{Q_{L}^{0}}\ \Omega _{2}\tilde{
\Phi}_{2}u_{R}^{0} -\overline{L_{L}^{0}}\ \Pi _{1}\Phi _{1}l_{R}^{0}-\overline{L_{L}^{0}}\
\Pi _{2}\Phi _{2}l_{R}^{0} + \text{h.c.}\ 
\label{Yuk}
\end{equation}
where $\Phi _{i}$ denote the Higgs doublets and $\widetilde{\Phi }_{i}\equiv
i\,\tau _{2}\,\Phi _{i}^{\ast }$ and $\Gamma _{i}$, $\Omega _{i}$, $\Pi_{i}$ are
matrices in flavour space. After spontaneous symmetry breaking, the
following quark mass matrices are generated: 
\begin{equation}
M_{d}=\frac{1}{\sqrt{2}}(v_{1}\Gamma _{1}+v_{2}e^{i\alpha }\Gamma _{2})\
,\qquad M_{u}=\frac{1}{\sqrt{2}}(v_{1}\Omega _{1}+v_{2}e^{-i\alpha }\Omega
_{2})  \label{mumd}
\end{equation}
where $v_{i}/\sqrt{2}\equiv |<0|\phi _{i}^{0}|0>|$ and $\alpha $ denotes the
relative phase of the two vacuum expectation values (vevs) of the neutral
components $\phi _{i}^{0}$ of $\Phi _{i}$. The neutral and the charged
Higgs interactions with quarks are of the form: 
\begin{eqnarray}
{\mathcal{L}}_{Y}(\mbox{quark, Higgs}) &=&-\overline{d_{L}^{0}}\,\frac{1}{v}%
\,[M_{d}\,H^{0}+N_{d}^{0}\,R+i\,N_{d}^{0}\,I]\,d_{R}^{0}+  \notag \\
&-&\overline{{u}_{L}^{0}}\,\frac{1}{v}\,[M_{u}\,H^{0}+N_{u}^{0}\,R+i%
\,N_{u}^{0}\,I]\,u_{R}^{0}+  \label{rep} \\
&&+\frac{\sqrt{2}H^{+}}{v}(\overline{{u}_{L}^{0}}N_{d}^{0}\,d_{R}^{0}-%
\overline{{u}_{R}^{0}}{N_{u}^{0}}^{\dagger }\,d_{L}^{0})+\text{h.c.}  \notag
\end{eqnarray}%
where $v\equiv \sqrt{v_{1}^{2}+v_{2}^{2}}\approx 246\ GeV$ and $H^{0}$, $R$
are orthogonal combinations of the fields $\rho _{j}$, arising when one
expands \cite{Lee:1973iz} the neutral scalar fields around their vevs, $\phi
_{j}^{0}=\frac{e^{i\alpha _{j}}}{\sqrt{2}}(v_{j}+\rho _{j}+i\eta _{j})$,
choosing $H^{0}$ in such a way that it has couplings to the quarks which are
proportional to the mass matrices, as can be seen from Eq.~(\ref{rep});
similarly, $I$ denotes the linear combination of $\eta _{j}$ orthogonal to
the neutral Goldstone boson. The matrices $N_{d}^{0}$ and $N_{u}^{0}$ are
given by: 
\begin{equation}
N_{d}^{0}=\frac{1}{\sqrt{2}}(v_{2}\Gamma _{1}-v_{1}e^{i\alpha }\Gamma
_{2}),\quad N_{u}^{0}=\frac{1}{\sqrt{2}}(v_{2}\Omega _{1}-v_{1}e^{-i\alpha
}\Omega _{2})  \label{ndnu}
\end{equation}
The quark mass matrices are diagonalized through

\begin{equation}
\begin{array}{c}
U_{dL}^{\dagger }\,M_{d}\,U_{dR}=D_{d}\equiv \mbox{diag}(m_{d},m_{s},m_{b}),
\\ 
\\ 
U_{uL}^{\dagger }\,M_{u}\,U_{uR}=D_{u}\equiv \mbox{diag}(m_{u},m_{c},m_{t}).
\end{array}
\label{mdmu}
\end{equation}
and the matrices $N_{d}^{0}$ and $N_{u}^{0}$ in the mass eigenstate basis
transform into: 
\begin{equation}
U_{dL}^{\dagger }\,N_{d}^{0}\,U_{dR}=N_{d},\qquad U_{uL}^{\dagger
}\,N_{u}^{0}\,U_{uR}=N_{u}.  \label{ndnu1}
\end{equation}
There are similar expressions for the leptonic sector. We do not introduce neutrino masses 
since these are not relevant for our analysis.

\section{Stability of the aligned 2HDM under RGE}

The aligned two Higgs doublet model (A2HDM) is defined  at tree level by the following relations involving the matrices introduced in Eq.~(\ref{Yuk}): 
\begin{eqnarray}
\Gamma _{2} &=&d\cdot \Gamma _{1}  \notag \\
\Omega _{2} &=&u\cdot \Omega _{1}  \label{AlignementConditions} \\
\Pi _{2} &=&e\cdot \Pi _{1}  \notag
\end{eqnarray}
where $d$, $u$, $e$ are constants.
In this section we analyse the stability of the A2HDM under renormalisation group 
equations (RGE).  The one loop renormalization group equations (RGE) for the Yukawa
couplings are \cite{Grimus:2004yh}, \cite{Ferreira:2010xe}: 
\begin{eqnarray}
\mathcal{D}\Gamma _{k} &=&a_{\Gamma }\Gamma _{k}+  \nonumber \\
&&+\sum_{l=1}^{2}\left[ 3\text{Tr}\!\left( \Gamma _{k}\Gamma _{l}^{\dagger }+\Omega
_{k}^{\dagger }\Omega _{l}\right) +\text{Tr}\!\left( \Pi _{k}\Pi _{l}^{\dagger
}+\Sigma _{k}^{\dagger }\Sigma _{l}\right) \right] \Gamma _{l}+  \nonumber \\
&&+\sum_{l=1}^{2}\left( -2\Omega _{l}\Omega _{k}^{\dagger }\Gamma
_{l}+\Gamma _{k}\Gamma _{l}^{\dagger }\Gamma _{l}+\frac{1}{2}\Omega
_{l}\Omega _{l}^{\dagger }\Gamma _{k}+\frac{1}{2}\Gamma _{l}\Gamma
_{l}^{\dagger }\Gamma _{k}\right) \ ,  \label{RGE1}
\end{eqnarray}
\begin{eqnarray}
\mathcal{D}\Omega _{k} &=&a_{\Omega }\Omega _{k}+  \nonumber \\
&&+\sum_{l=1}^{2}\left[ 3\text{Tr}\!\left( \Omega _{k}\Omega _{l}^{\dagger }+\Gamma
_{k}^{\dagger }\Gamma _{l}\right) +\text{Tr}\!\left( \Sigma _{k}\Sigma _{l}^{\dagger
}+\Pi _{k}^{\dagger }\Pi _{l}\right) \right] \Omega _{l}+  \nonumber \\
&&+\sum_{l=1}^{2}\left( -2\Gamma _{l}\Gamma _{k}^{\dagger }\Omega
_{l}+\Omega _{k}\Omega _{l}^{\dagger }\Omega _{l}+\frac{1}{2}\Gamma
_{l}\Gamma _{l}^{\dagger }\Omega _{k}+\frac{1}{2}\Omega _{l}\Omega
_{l}^{\dagger }\Omega _{k}\right) \ ,  \label{RGE2}
\end{eqnarray}
\begin{eqnarray}
\mathcal{D}\Pi _{k} &=&a_{\Pi }\Pi _{k}+  \nonumber \\
&&+\sum_{l=1}^{2}\left[ 3\text{Tr}\!\left( \Gamma _{k}\Gamma _{l}^{\dagger }+\Omega
_{k}^{\dagger }\Omega _{l}\right) +\text{Tr}\!\left( \Pi _{k}\Pi _{l}^{\dagger
}+\Sigma _{k}^{\dagger }\Sigma _{l}\right) \right] \Pi _{l}+  \nonumber \\
&&+\sum_{l=1}^{2}\left( -2\Sigma _{l}\Sigma _{k}^{\dagger }\Pi _{l}+\Pi
_{k}\Pi _{l}^{\dagger }\Pi _{l}+\frac{1}{2}\Sigma _{l}\Sigma _{l}^{\dagger
}\Pi _{k}+\frac{1}{2}\Pi _{l}\Pi _{l}^{\dagger }\Pi _{k}\right) \ ,
\label{RGE3}
\end{eqnarray}
where $\mathcal{D}\equiv 16\pi ^{2}\mu \left( d/d\mu \right) $ and $\mu $ is
the renormalization scale.  The coefficients 
$a_{\Gamma }$, $a_{\Omega }$ and $a_{\Pi }$ are given by:
\begin{eqnarray} 
 a_{\Gamma }= -8 g_s^2 -\frac{9}{4} g^2 -\frac{5}{12} {g^\prime}^2 \\
a_{\Omega }=  -8 g_s^2 -\frac{9}{4} g^2 -\frac{17}{12} {g^\prime}^2  \\
a_{\Pi } =   -\frac{9}{4} g^2 -\frac{15}{4} {g^\prime}^2
\end{eqnarray}
where $g_s$, $g$ and $g^\prime$ are the gauge coupling constants of
$SU(3)_c$, $SU(2)_L$ and $U(1)_Y$ respectively. The alignment relations
given by Eq.~(\ref{AlignementConditions})  guarantee the absence of Higgs
mediated FCNC at tree level because both matrices $M_{d}$ and $N_{d}$ 
are proportional to $\Gamma _{1}$. Similarly both
$M_{u}$ and $N_{u}$ are proportional to $\Omega _{1}$ and 
$M_{l}$, $N_{l}$ to  $\Pi_{1}$. In general, these relations are
broken at one loop level. From Eqs.~(\ref{RGE1}),(\ref{RGE2}),(\ref{RGE3}) one can easily derive:
\begin{eqnarray}
\mathcal{D}\left( \Gamma _{2}\right) -d\cdot \mathcal{D}\left( \Gamma
_{1}\right) &=&\left( u^{\ast }-d\right) \left( 1+ud\right) \left\{
3Tr\left( \Omega _{1}^{\dagger }\Omega _{1}\right) -2\Omega _{1}\Omega
_{1}^{\dagger }\right\} \Gamma _{1}+  \notag \\
&&+\left( e-d\right) \left( 1+e^{\ast }d\right) Tr\left( \Pi _{1}^{\dagger
}\Pi _{1}\right) \Gamma _{1}  \label{AlignmentBrokenDown}
\end{eqnarray}
\begin{eqnarray}
\mathcal{D}\left( \Omega _{2}\right) -u\cdot \mathcal{D}\left( \Omega
_{1}\right) &=&\left( d^{\ast }-u\right) \left( 1+ud\right) \left\{
3Tr\left( \Gamma _{1}^{\dagger }\Gamma _{1}\right) -2\Gamma _{1}\Gamma
_{1}^{\dagger }\right\} \Omega _{1}+  \notag \\
&&+\left( e^{\ast }-u\right) \left( 1+eu\right) Tr\left( \Pi _{1}^{\dagger
}\Pi _{1}\right) \Omega _{1}  \label{AlignmentBrokenUp}
\end{eqnarray}
\begin{eqnarray}
\mathcal{D}\left( \Pi _{2}\right) -e\cdot \mathcal{D}\left( \Pi _{1}\right)
 &=& 3\left( d-e\right) \left( 1+d^{\ast }e\right) Tr\left( \Gamma _{1}^{\dagger
}\Gamma _{1}\right)  \Pi _{1} +  \notag \\  
&&+ 3\left( u^{\ast }-e\right) \left( 1+eu\right) Tr\left( \Omega _{1}^{\dagger
}\Omega _{1}\right) 
 \Pi _{1}  \label{AlignmentBrokenLepton}
\end{eqnarray}
In order to enforce Eq.~(\ref{AlignementConditions}) at one loop level it is
easy to realize that it is sufficient to impose: 
\begin{eqnarray}
\mathcal{D}\left( \Gamma _{2}\right) -d\cdot \mathcal{D}\left( \Gamma
_{1}\right)  &\propto &\Gamma _{1}  \label{RGEstableDown} \\
\mathcal{D}\left( \Omega _{2}\right) -u\cdot \mathcal{D}\left( \Omega
_{1}\right)  &\propto &\Omega _{1}  \label{RGEstableUp} \\
\mathcal{D}\left( \Pi _{2}\right) -e\cdot \mathcal{D}\left( \Pi _{1}\right) 
&\propto &\Pi _{1}  \label{RGEstableLepton}
\end{eqnarray}
in fact the proportionality constants on the r.h.s. are the running \footnote
{The autors of reference  \cite{Ferreira:2010xe} impose the r.h.s. \ of Eqs.~(\ref
{AlignmentBrokenDown}),(\ref{AlignmentBrokenUp},(\ref{AlignmentBrokenLepton})
to be equal to zero. This amounts to imposing alignment at one loop level and
imposing additionally that there is no running of the parameters $u,d$ and $e$.}
of $d,u$ and $e $.
Therefore Eq.~(\ref{AlignmentBrokenLepton}) does not impose any constraint: at
one loop level the charged lepton sector remains aligned and there are no
FCNC in the leptonic sector. This result agrees with the findings of
references \cite{Braeuninger:2010td} and \cite{Jung:2010ik}.

In equations (\ref{AlignmentBrokenDown}), and (\ref{AlignmentBrokenUp}) the pieces
that can break the alignment in the quark sector are the terms: $\Omega
_{1}\Omega _{1}^{\dagger }\Gamma _{1}$ and $\Gamma _{1}\Gamma _{1}^{\dagger
}\Omega _{1}$ respectively$\footnote
{It can be readily seen that $\Omega _{1}\Omega _{1}^{\dagger }\Gamma
_{1}\propto M_{u}M_{u}^{\dag }M_{d}$ and $\Gamma _{1}\Gamma _{1}^{\dagger
}\Omega _{1}\propto M_{d}M_{d}^{\dag }M_{u}$. It is worth emphasizing
that these
structures are precisely the ones obtained in  \cite{Braeuninger:2010td}  and in 
\cite{Jung:2010ik},
which produce FCNC at one loop level.}$. In order to have
alignment at one loop level - fulfilling  Eqs(\ref{RGEstableDown}), (\ref
{RGEstableUp})- there are two types of solutions:

\begin{enumerate}
\item $\left( u^{\ast }-d\right) \left( 1+ud\right) =0$

\item $\Omega _{1}\Omega _{1}^{\dagger }\Gamma _{1}=\lambda _{\Gamma }\Gamma
_{1}$ and $\Gamma _{1}\Gamma _{1}^{\dagger }\Omega _{1}=\lambda _{\Omega
}\Omega _{1}$. With $\lambda _{\Gamma }$ and $\lambda _{\Omega }$ complex
numbers.
\end{enumerate}
Solutions of type 1 include the usual 2HDM with natural flavour conservation (NFC), 
where the up and down quarks receive contributions from only one Higgs doublet. It
is well known that this can be achieved through the introduction of a $Z_{2}$ symmetry.
Here, we are not interested in this class of  well known solutions. We are 
interested in the class of solutions of type 2, and in the Appendix, we have studied
the complete set of matrices $\Omega _{1}$ and $\Gamma _{1}$ that
obey to the conditions required for this class of solutions.  We have shown in
the Appendix, that if one requires stability under the RGE and at the same time 
Yukawa structures which are, in leading order, in agreement with the observed
pattern of quark masses and mixing, then one is lead to a unique solution,
where the matrices $\Omega _{1}$ and $\Gamma _{1}$ are of the form:
\begin{equation}
\begin{array}{ccc}
\Omega _{1}=c_{1}^{d}\Delta & ; & \Gamma _{1}=c_{1}^{u}\Delta
\label{equ20}
\end{array}
\end{equation}
with $\Delta $  the democratic mass matrix:
\begin{equation}
\Delta =\left( 
\begin{array}{ccc}
1 & 1 & 1 \\ 
1 & 1 & 1 \\ 
1 & 1 & 1
\end{array}
\right)
\end{equation}
This solution corresponds to the limit where only the top and bottom quarks acquire
mass, while the two first generations are massless. The up and down quarks are 
aligned in flavour space, so the $V_{CKM}$ matrix equals the identity. The other
stable solutions of type 2 correspond to  non realistic cases like for example  
having all up or down quarks massless or two up or two down quark masses 
degenerate or with a  $V_{CKM}$ very far from the identity matrix.

It is remarkable to realise that the so called democratic mass matrix is
stable under RGE and that precisely this stability also enforces what could
be called "quark alignment" in the sense that we also have a proportionality
among $\Gamma _{i}$ and $\Omega _{i}$.

\section{Natural Quasi-Alignment of Yukawa couplings}

In this section we search for the minimal symmetry which when imposed on the
Lagrangian, leads to the stable solution described in the previous section,
corresponding to the democratic Yukawa couplings of   Eq.~(\ref{equ20}). Before 
describing this symmetry, it is worth analysing another type of alignment
which is verified experimentally, the so-called up-down alignment in the
quark sector.


\subsection{The up-down alignment in the quark sector}

In the quark sector, flavour mixing is small. This means that there is a
weak basis (WB) where both $M_u$, and $M_d$ are close to the diagonal form.
Experiment indicates that not only flavour mixing is small, but there is
also up-down flavour alignment in the quark sector in the following sense.
We can choose, without loss of generality, a WB where $M_u = \mbox{diag}
(m_u, m_c, m_t)$. Of course, this is just a choice of ordering, with no
physical meaning. Small mixing implies that in this WB $M_d$ is almost
diagonal. In principle, since the Yukawa couplings $Y_u$, $Y_d$ are not
constrained in the SM, there is equal probability of $M_d$ being close to $%
M_d = \mbox{diag} (m_d, m_s, m_b)$. corresponding to up-down alignment, or
being close for instance to $M_d = \mbox{diag} (m_b, m_s, m_d)$ in which
case there is up-down misalignment. It is clear that in the SM, assuming
small mixing and hierarchical quark masses, the probability of obtaining
up-down alignment is only 1/6. Given a set of arbitrary quark mass
matrices $M_u$, $M_d$, one can derive necessary and sufficient conditions to
obtain small mixing and up-down alignment, expressed in terms of WB
invariants \cite{Branco:2011aa}.
Since the experimentally verified up-down alignment is not
automatic in the SM, one may wonder whether there is a symmetry which leads
to up-down alignment. In the next subsection, we propose a symmetry which
leads to up-down alignment in the quark sector and when extended to a 2HDM
leads to a natural alignment of the two Higgs doublets in flavour space.

\subsection{$Z_3 \times Z^\prime_3$ symmetry and the two Higgs alignment}

We introduce the following $Z_{3}\times Z_{3}^{\prime }$ symmetry under
which the quark left-handed doublets $Q_{L_{i}}^{0}$ , the right-handed up
quarks $u_{L_{i}}^{0}$, and the right-handed down quarks $d_{L_{i}}^{0}$
transform in the following way: 
\begin{equation}
\begin{aligned} Q^0_{L_i} &\longrightarrow P^{\dagger}_{ij}\,Q^0_{L_j},\\
u^0_{R_i} &\longrightarrow P_{ij}\,u^0_{R_j},\\ d^0_{R_i} &\longrightarrow
P_{ij}\,d^0_{R_j}, \end{aligned} \label{222}
\end{equation}
where $Z_{3}$ corresponds to $P=\mathbbm{1}+E_{1}$ and $Z_{3}^{\prime }$ to $
P=\mathbbm{1}+E_{2}$ with: 
\begin{equation}
E_{1}=\frac{\omega -1}{2}\left( 
\begin{array}{ccc}
1 & -1 & 0 \\ 
-1 & 1 & 0 \\ 
0 & 0 & 0
\end{array}
\right) ;\qquad E_{2}=\frac{\omega -1}{3}\left( 
\begin{array}{ccc}
\frac{1}{2} & \frac{1}{2} & -1 \\ 
\frac{1}{2} & \frac{1}{2} & -1 \\ 
-1 & -1 & 2
\end{array}
\right)  \label{e1e2}
\end{equation}
and $\omega =e^{\frac{2\pi i}{3}}$. The Higgs doublets transform trivially
under $Z_{3}\times Z_{3}^{\prime }$. The above symmetry leads to the
following form for the Yukawa matrices $\Gamma _{j}= c_{j}^{d}\Delta $,
$\Omega _{j} =c_{j}^{u}\Delta$, corresponding to the stable solution 
of Eq.~(\ref{equ20}).
This can easily be checked since $\Delta E_{1}=\Delta E_{2}=0$. 
We thus conclude that the symmetry of Eqs.~(\ref{222}), (\ref{e1e2})
leads to the alignment of the two Yukawa coupling matrices, with a democratic
flavour structure. Note that this solution also guarantees an up-down alignment
in the quark sector, as defined in the previous subsection. \\
In order to give mass to the first two quark generations, the 
$Z_{3}\times Z_{3}^{\prime }$ symmetry has to be broken. This breaking
will also lead to Higgs mediated FCNC, but these couplings 
will be suppressed by the smallness of the quark masses. In order to illustrate 
how a realistic pattern of quark masses and mixing can be obtained, we 
shall assume that the breaking of the   $Z_{3}\times Z_{3}^{\prime }$ symmetry
occurs in two steps. In the first step the symmetry $Z_{3}\times Z_{3}^{\prime }$
is broken into just one of the $Z_{3}$ and the second generation acquires
mass and finally in the last step the mass of the quarks $u$, $d$ is
generated.
In the first step the symmetry $Z_{3}\times Z_{3}^{\prime }$ is broken to $
Z_{3}$ generated by $P=\mathbbm{1}+E_{1}$. One can check that: 
\begin{equation}
\Gamma _{j}=c_{j}^{d}\left( \Delta +\varepsilon _{d}\,A\right) ;\qquad
\Omega _{j}=c_{j}^{u}\left( \Delta +\varepsilon _{u}\,A\right) ;\qquad
A=\left( 
\begin{array}{ccc}
0 & 0 & 1 \\ 
0 & 0 & 1 \\ 
1 & 1 & 1
\end{array}
\right)   \label{go}
\end{equation}
are invariant under this $Z_{3}$ symmetry. Note that $A\,E_{1}=0$. At this
stage the second generation acquires mass. Finally, the lightest quarks, $u$
and $d$ acquire mass through a small perturbation, proportional to $\hat{
\delta}_{d,u}$, which breaks this $Z_{3}$ symmetry. We assume that: 
\begin{equation}
\Gamma _{2}=c_{2}^{d}\left( \Delta +\varepsilon _{d}\,A+\hat{\delta}%
_{d}\,B_{d}\right)   \label{g1}
\end{equation}
while 
\begin{equation}
\Gamma _{1}=c_{1}^{d}\left( \Delta +\varepsilon _{d}\,A\right) ;  \label{g2}
\end{equation}
with equivalent expressions for the up sector. Here
\begin{equation}
B_{u}=\left( 
\begin{array}{ccc}
0 & 0 & 1 \\ 
0 & 0 & 0 \\ 
1 & 0 & 1
\end{array}
\right);\qquad B_{d}=\left( 
\begin{array}{ccc}
0 & 0 & 1 \\ 
0 & 0 & 0 \\ 
1 & 0 & \eta 
\end{array}
\right)   \label{bud}
\end{equation}
where $\eta $ is some complex number with modulus of order one. The symmetry
is broken, and neither $B_{u}$ nor $B_{d}$ are invariant under the $
Z_{3}\times Z_{3}^{\prime }$ symmetry.

\section{Suppression of scalar mediated FCNC}
In order to study the suppression of scalar mediated FCNC, it is useful to start
by analysing the parameter space in our framework.

\subsection{The Parameter Space}

From Eqs. (\ref{mumd}), (\ref{g1}) and (\ref{g2}) it follows that, in leading order 
\begin{equation}
m_{b}=\frac{3}{\sqrt{2}}\left\vert c_{1}^{d}v_{1}+c_{2}^{d}v_{2}e^{i\alpha
}\right\vert \quad ;\quad m_{t}=\frac{3}{\sqrt{2}}\left\vert
c_{1}^{u}v_{1}+c_{2}^{u}v_{2}e^{-i\alpha }\right\vert  \label{conv}
\end{equation}
Writing $v\equiv \sqrt{v_{1}^{2}+v_{2}^{2}}=v_{1}\sqrt{1+t^{2}}$, with
\begin{equation}
t=\frac{v_{2}}{v_{1}}  \label{t}
\end{equation}
we obtain in leading order the following relations:
\begin{equation}
\frac{\left\vert c_{1}^{d}+c_{2}^{d}\ te^{i\alpha }\right\vert }{\sqrt{
1+t^{2}}}=\frac{\sqrt{2}}{3}\frac{m_{b}}{v}\quad ;\quad \frac{\left\vert
c_{1}^{u}+c_{2}^{u}\ te^{-i\alpha }\right\vert }{\sqrt{1+t^{2}}}=\frac{\sqrt{%
2}}{3}\frac{m_{t}}{v}  \label{ci}
\end{equation}
which impose restrictions on the allowed parameter space.
A priori, we do not assume any conspiracy between parameters and take $
t=O(1)$. 
It is then clear from Eq. (\ref{ci}), that the $c_{i}^{u}$ are generically
of order one, while $c_{i}^{d}$ are smaller, and may assume values of order $
O(\frac{m_{b}}{m_{t}})$. This is an important ingredient which, as we shall
see, will play a r\^{o}le in the evaluation of the strength of FCNC's and the allowed
parameter space for the  Higgs masses.

Next we give the structure of the flavour changing neutral Yukawa couplings.
For that,  it is useful
to express the quark mass matrices in Eq. (\ref{mumd}) in terms of the
perturbations given in Eqs. (\ref{g1}), (\ref{g2}):
\begin{equation}
\begin{array}{l}
M_{d}=\frac{v_{1}}{\sqrt{2}}\left( c_{1}^{d}+c_{2}^{d}te^{i\alpha }\right) \ 
\left[ \Delta +\varepsilon _{d}\ A+\delta _{d}\ B_{d}\right] \\ 
\\ 
M_{u}=\frac{v_{1}}{\sqrt{2}}\left( c_{1}^{u}+c_{2}^{u}te^{-i\alpha }\right)
\ \left[ \Delta +\varepsilon _{u}\ A+\delta _{u}\ B_{u}\right]%
\end{array}
\quad ;\quad 
\begin{array}{c}
\delta _{d}\equiv \frac{c_{2}^{d}te^{i\alpha }}{c_{1}^{d}+c_{2}^{d}te^{i%
\alpha }}\ \widehat{\delta }_{d} \\ 
\\ 
\delta _{u}\equiv \frac{c_{2}^{u}te^{-i\alpha }}{c_{1}^{u}+c_{2}^{u}te^{-i%
\alpha }}\ \widehat{\delta }_{u}
\end{array}
\label{quark1}
\end{equation}

Then, we derive the expressions for the matrices which
couple to the Higgs scalars in Eqs. (\ref{rep}), (\ref{ndnu}). In the basis
where the up and down quark matrices are diagonal, the matrices $N_{d}$, $
N_{u}$ of Eq. ( \ref{ndnu1}) become 
\begin{equation}
\begin{array}{c}
N_{d}=t\ D_{d}-\frac{v_{1}}{\sqrt{2}}\left( 1+t^{2}\right) e^{i\alpha }\
U_{d_{L}}^{\dagger }\ \Gamma _{2}\ U_{d_{R}} \\ 
\\ 
N_{u}=t\ D_{u}-\frac{v_{1}}{\sqrt{2}}\left( 1+t^{2}\right) e^{-i\alpha }\
U_{u_{L}}^{\dagger }\ \Omega _{2}\ U_{u_{R}}
\end{array}
\label{ns}
\end{equation}
where we have used Eqs.~(\ref{ndnu}) and (\ref{ndnu1}) with Eq.~(\ref{mumd}).
Finally, from Eq.~(\ref{ns}) combined with Eqs.~(\ref{g1}) and Eq. (\ref{quark1})
we find 
\begin{equation}
\begin{array}{l}
N_{d}=\ \frac{c_{1}^{d}t-c_{2}^{d}e^{i\alpha }}{
c_{1}^{d}+c_{2}^{d}te^{i\alpha }}\ D_{d} - {\frac{v}{\sqrt{2}}}\ 
\frac{c_{1}^{d}\sqrt{1+t^{2}}}{t}\ \delta _{d}\ 
U_{d_{L}}^{\dagger }\ B_{d}\ U_{d_{R}} \\ 
\\ 
N_{u}=\frac{c_{1}^{u}t-c_{2}^{u}e^{-i\alpha }}{%
c_{1}^{u}+c_{2}^{u}te^{-i\alpha }}\ D_{u} - {\frac{v}{\sqrt{2}}} \frac{c_{1}^{u}\sqrt{%
1+t^{2}}}{t}\ \delta _{u}\ U_{u_{L}}^{\dagger }\ B_{u}\ U_{u_{R}}%
\end{array}
\label{ns1}
\end{equation}
where $D_{d}\equiv diag(m_{d},m_{s},m_{b})$, $D_{u}\equiv
diag(m_{u},m_{c},m_{t})$.

The crucial point is that in our scheme  these matrices have an extra
suppression factor, proportional to $\delta _{d,u}$.
Using the expressions given in Eqs. (\ref{quark1}), (\ref{bud}) and
computing the trace, second invariant and determinant for the squared quark
mass matrices $H_{u,d}\equiv \left( MM^{\dagger }\right) _{u,d}$ , one can
find that in leading order:
\begin{equation}
\begin{array}{c}
\delta _{d}=\sqrt{3}\sqrt{\frac{m_{d}}{m_{s}}}\frac{m_{s}}{m_{b}}=O\left(
\lambda ^{3}\right)  \\ 
\\ 
\delta _{u}=\sqrt{3}\sqrt{\frac{m_{u}}{m_{c}}}\frac{m_{c}}{m_{t}}=O\left(
\lambda ^{5}\right) 
\end{array}
\label{dud}
\end{equation}
where $\lambda \equiv 0.2$ is of the order of the Cabibbo angle.

From Eq. (\ref{quark1}) it follows  that in leading order $
U_{d_{L}}=U_{u_{L}}=U_{d_{R}}=U_{u_{R}}=F$, where
\begin{equation}
F==\left( 
\begin{array}{ccc}
1/\sqrt{2} & 1/\sqrt{6} & 1/\sqrt{3} \\ 
-1/\sqrt{2} & 1/\sqrt{6} & 1/\sqrt{3} \\ 
0 & -2/\sqrt{6} & 1/\sqrt{3}%
\end{array}
\right)   \label{f}
\end{equation}
is the matrix that diagonalizes the exact democratic limit $\Delta $. Thus,
taking into account  Eq. (\ref{bud}), the matrix contributions from $U_{d_{L}}^{\dagger }\
B_{d}\ U_{d_{R}}$ and $U_{u_{L}}^{\dagger }\ B_{u}\ U_{u_{R}}$ are both of
order one. One can thus conclude that:

--for the down sector, with the assumptions made after Eq. (\ref{ci}), we
have a total suppression factor of $O(\frac{m_{b}}{m_{t}})\cdot O\left(
\lambda ^{3}\right) $

--for the up sector, we have a suppression factor of $O\left( \lambda
^{5}\right) $ or smaller depending on the value that we choose to assume for 
$c_{1}^{u}$, but which, as explained, it is reasonable to take of order one.

\section{Numerical analysis}

The matrices of Eq.~\eqref{quark1} may be explicitly written as
\begin{equation}
M_{u}= c_{u} \begin{pmatrix}
1 & 1 & 1 + \varepsilon + \delta\\
1 & 1 & 1 + \varepsilon\\
1 + \varepsilon + \delta & 1 + \varepsilon & 1 + \varepsilon + \delta
\end{pmatrix}_{u},\quad
M_{d}= c_{d} \begin{pmatrix}
1 & 1 & 1 + \varepsilon + \delta\\
1 & 1 & 1 + \varepsilon\\
1 + \varepsilon + \delta & 1 + \varepsilon & 1 + \varepsilon + \eta\,\delta
\end{pmatrix}_{d},
\end{equation}
where we have introduced $c_d\equiv\frac{v_1}{\sqrt{2}}(c^d_1+c^d_2\,t\,e^{i\alpha})$, and $c_u \equiv \frac{v_1}{\sqrt{2}}(c^u_1+c^u_2\,t\,e^{-i\alpha})$. Although these two coefficients are in general complex, and since the physically meaningful matrices are those defined as $H=M\,M^\dagger$, both coefficients
may be taken as real for our numerical exercise. If one then parametrizes the remaining variables as
\begin{equation}
\varepsilon = \varepsilon_m \exp\left(i\,\varepsilon_f\right),\qquad \delta = \delta_m \exp\left(i\,\delta_f\right),\qquad \eta = \eta_m \exp\left(i\,\eta_f\right),
\end{equation}
one is left with twelve real parameters that compose the quark mass matrices in our scheme.

In order to check if this parameter space could accommodate the flavour sector, a numerical survey was made where we looked for one combination that could fit the observed values of the quark masses given at the scale of the Z boson mass \cite{Antusch:2013jca}, the moduli of the entries of the CKM matrix \cite{ckmfitter}, 
the strength of CP violation $I_{CP}$ and $\sin 2\beta$ and $\gamma$ \cite{ckmfitter}, with $\beta$ and $\gamma$ being two of the angles of the unitarity triangle. A simple run of all twelve
parameters produced a ``reference point'':
\par
\begin{center}
\begin{tabular}{ccc}
 \hline\hline
 & Up-sector & Down-sector\\
 \hline
 $c$ & 56.73 & 0.89\\
 $\varepsilon_m$ & $1.6 \times 10^{-2}$ & 0.11\\
 $\varepsilon_f$ &$ -5.6 \times 10^{-3}$ & 0.41\\
 $\delta_m$ & $8.1 \times 10^{-4}$ &$ 2.2 \times 10^{-2}$\\
 $\delta_f$ &$ \pi + 0.32$ & 2.26\\
 $\eta_m$ & --- & 4.99\\
 $\eta_f$ & --- & $\pi + 0.62$ \\ \hline \hline
\end{tabular}
\end{center}
\par
\noindent which yields the output values:
\begin{equation}
\begin{aligned}
D_d &= \textrm{diag}(0.00204,0.05824,2.85356) \ \mbox{GeV},\\
D_u &= \textrm{diag}(0.00114,0.61736,171.684) \ \mbox{GeV},\\
| V_{CKM}| &= \begin{pmatrix}
0.9745 & 0.2244 & 0.0036\\
0.2243 & 0.9737 & 0.0415\\
0.0087 & 0.0407 & 0.9991
\end{pmatrix},\\
|I_{CP}| &= 3.0 \times 10^{-5},\\
\sin 2\beta &= 0.69,\\
\gamma &= 69.3^\circ.
\end{aligned}
\end{equation}

\begin{figure}
\centering
\subfigure{
  \centering
  \includegraphics[width = 0.427\textwidth]{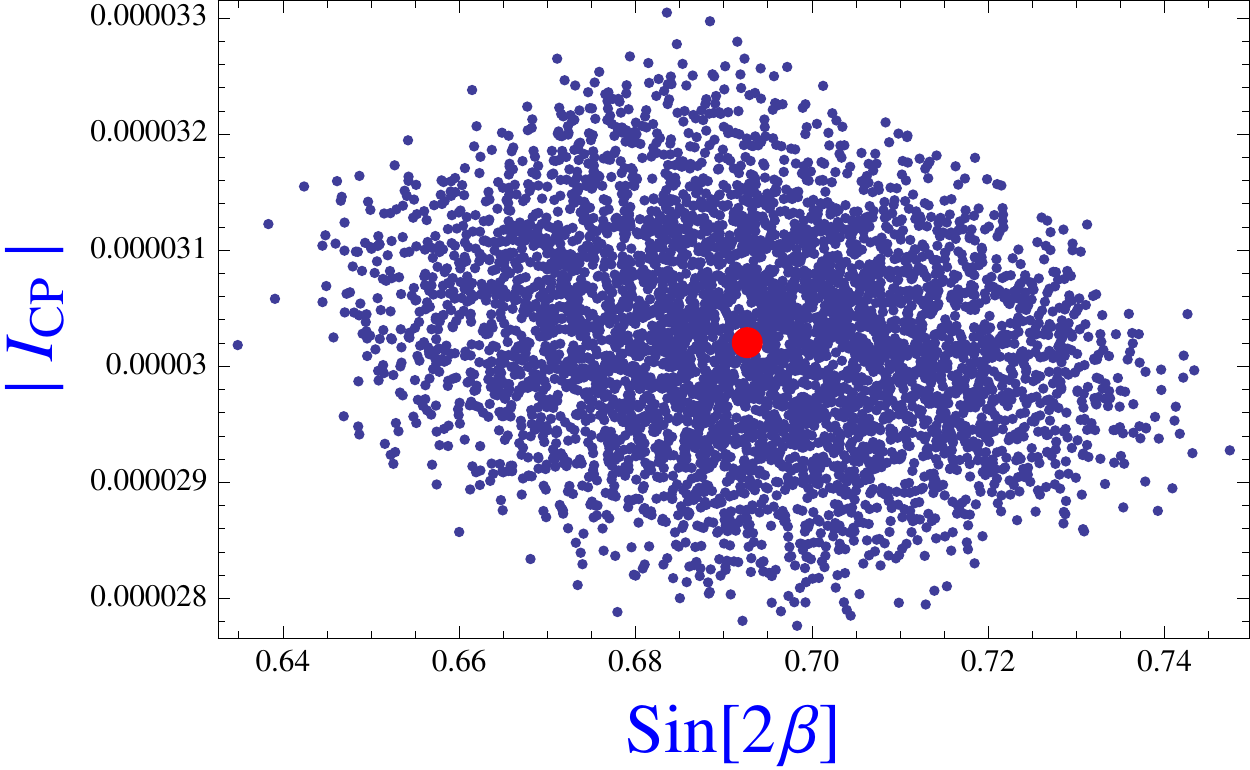}}\quad
\subfigure{
  \centering
  \includegraphics[width = 0.4\textwidth]{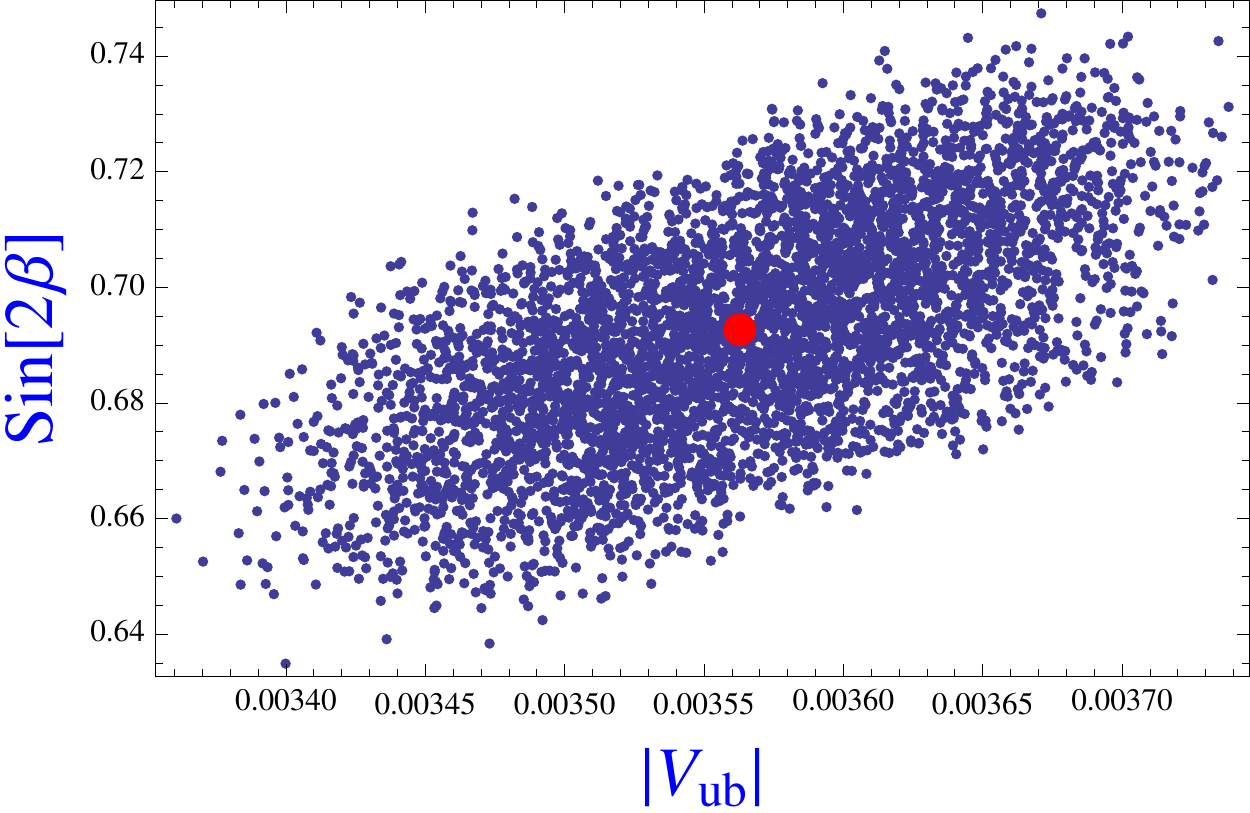}}
\caption{We present scatter plots showing $|I_{CP}|$ versus $\sin 2 \beta$ 
and $\sin 2 \beta$ versus $|V_{ub}|$ obtained by varying randomly the
input parameters around the reference point.}
\label{fig:scatter}
\end{figure}

It should be noted that the twelve parameters fix not only $V_{CKM}$
and the quark mass spectrum, but also the strength of all the FCNC couplings.
In order to evaluate the numerical stability of this reference point, we performed 
a numerical check, varying the input parameters randomly around the values 
that produced the reference point above; 
the new results were then combined in the scatter plots shown in 
Fig.~\ref{fig:scatter} where the reference point is highlighted. 
In order to obtain an estimate of the lower bound for the flavour-violating
Higgs masses, we consider the contribution to $K^0-\overline{K}^0$ mixing. 
Apart from the SM box diagram
one now has a New Physics contribution arising from
the scalar-mediated FCNC tree-level diagrams thus making the total transition amplitude equal to $M_{12} = M_{12}^{\textrm{SM}} + M_{12}^{\textrm{NP}}$, \cite{Botella:2014ska} with:
\begin{equation}
M_{12}^{NP} = \sum_{H=R,I} \frac{f_M^2\,m_M}{96\,v^2\,m_H^2}\,\bigg\{\bigg[1+\bigg(\frac{m_M}{m_{q1}+m_{q2}}\bigg)^2\bigg]C_1(H) - \bigg[1+11\bigg(\frac{m_M}{m_{q1}+m_{q2}}\bigg)^2\bigg]C_2(H)\bigg\},
\end{equation}
where: 
\begin{equation}
C_1(R) = \left[\left( N_{q2q1}\right)^{*} + N_{q1q2} \right]^2,\quad
C_2(R) = \left[\left( N_{q2q1}\right)^{*} - N_{q1q2} \right]^2,
\end{equation}
and
\begin{equation}
C_1(I) = -\left[\left( N_{q2q1}\right)^{*} - N_{q1q2} \right]^2,\quad
C_2(I) = -\left[\left( N_{q2q1}\right)^{*} + N_{q1q2} \right]^2.
\end{equation}
The indices $q_1$ and $q_2$ refer to the valence quarks of the meson $M$, and 
$N$ is $N_u$ or $N_d$ depending on the meson system considered.

\begin{figure}
\centering
  \includegraphics[width = 0.4\textwidth]{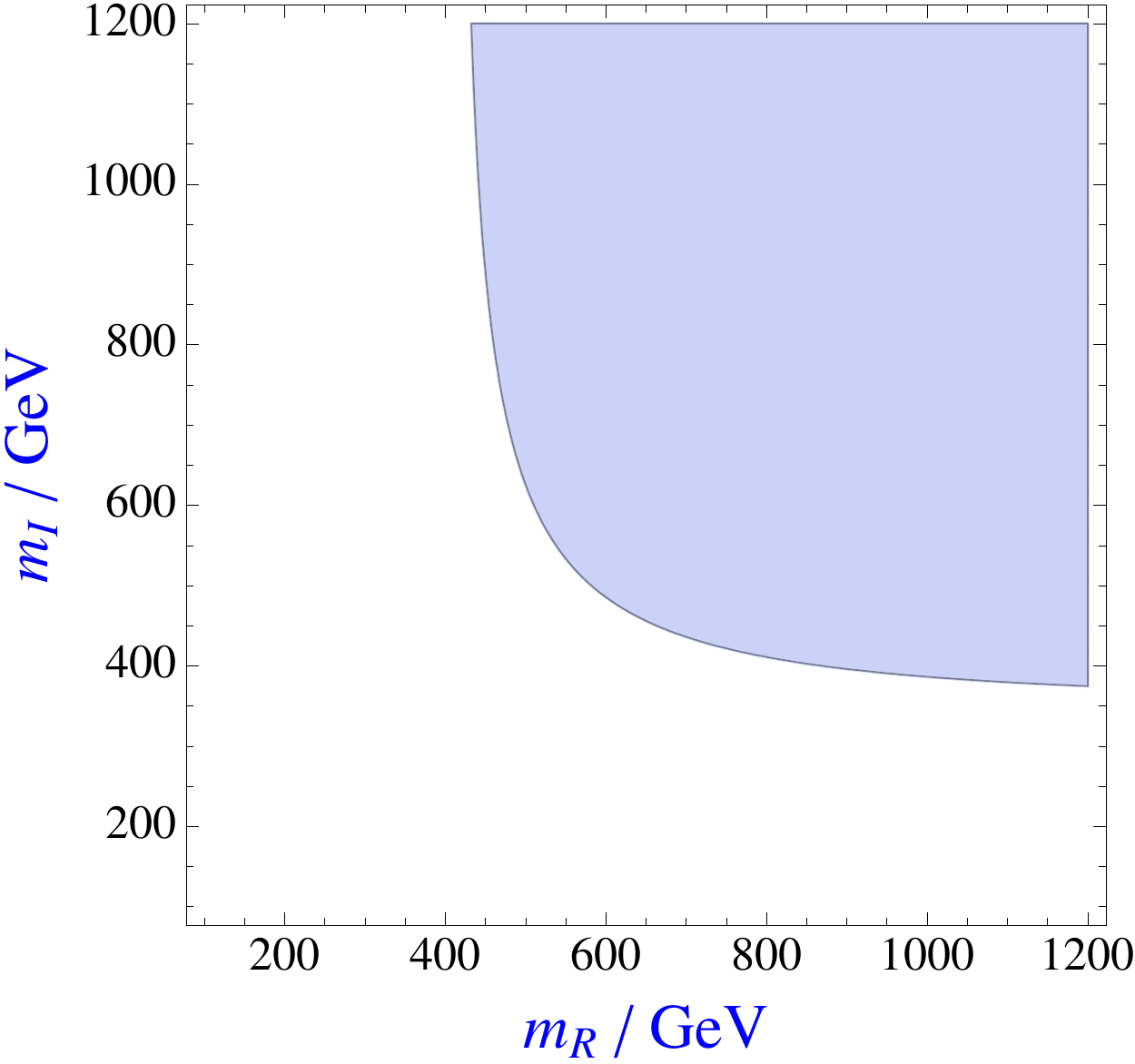}
\caption{Plot showing the allowed region for $m_I$ and $m_R$, taking 
into account the constraint on $\Delta m_K$.}
\label{fig:region}
\end{figure}

In this framework it is a good approximation to use the matrix $F$ for 
both $U_{d_L}$ and $U_{d_R}$. 
Using the values we obtained for $\delta_d$ and taking, as already discussed, $t\simeq 1$ and $c_1^d\simeq 
\frac{\sqrt{2}}{3}\frac{m_b}{v}$, the new physics contribution to $M_{12}^K$ 
becomes solely dependent on $f_K$, $m_K$, $m_R$ and $m_I$. In $K^0-
\overline{K}^0$, both $M_{12}^K$ and $\Gamma_{12}^K$ are relevant for the 
mass difference $\Delta m_K$. It is reasonable to impose the constraint 
that $M_{12}^{\textrm{NP}}$ in the neutral kaon system does not exceed the 
experimental value of $\Delta m_K$. Adopting as input values the PDG experimental determinations of $f_K$, $m_K$ and $\Delta m_K$ \cite{PDG:2014}, one is left with combinations of $m_R$ and $m_I$ where our model respects the inequality $M_{12}^{NP\,(K)}<\Delta m_K$. The region plot that we have obtained is presented in Fig.\ref{fig:region}. It is clear that in this framework, the masses of the flavour-violating
neutral Higgs can be  below the TeV scale, so that they could be 
discovered at the next run of the LHC.

\section{Conclusions}
We have studied in detail, in the framework of 2HDM, the question of stability 
of alignment, under the renormalization group. It was shown that there are new 
stable solutions, apart from those found in Ref.~\cite{Ferreira:2010xe}. Stability under the RGE puts very strict restrictions on the flavour structure of the Higgs
Yukawa couplings. If one imposes the stability conditions and at the same time 
requires that the flavour structure is in agreement with the observed pattern
of quark masses and mixing, then one is lead to a unique solution, where
all Higgs flavour matrices are proportional to the so-called democratic matrix. We 
have also shown that these flavour structures leading to stable alignment
can be obtained by imposing on the Lagrangian a $Z_{3}\times Z_{3}^{\prime }$
symmetry.
In the limit where this symmetry is exact, only the 
third generation of quarks acquires a mass. Non-vanishing masses for the two
first generations are obtained through the breaking of the discrete symmetry
which in turn generates scalar mediated FCNC which are suppressed by the
smallness of the light quark masses. The scenario presented in this paper
establishes a possible intriguing link between stability of alignment in 2HDM
and the observed pattern of quark masses and mixing.

\section*{Appendix: Solutions to the alignment conditions}

The solutions to the alignment conditions
\begin{equation}
\begin{array}{ccc}
\Omega _{1}\Omega _{1}^{\dagger }\Gamma _{1}=\lambda _{\Gamma }\Gamma _{1} & 
; & \Gamma _{1}\Gamma _{1}^{\dagger }\Omega _{1}=\lambda _{\Omega }\Omega
_{1}%
\end{array}
\label{A1}
\end{equation}
can be obtained by the following steps. First we define the Hermitian
matrices 
\begin{equation}
\begin{array}{ccc}
H_{\Gamma }=\Gamma _{1}\Gamma _{1}^{\dag } & ; & H_{\Omega }=\Omega
_{1}\Omega _{1}^{\dagger }
\end{array}
\label{A2}
\end{equation}
It is easy to show that $\lambda _{\Gamma }$ and $\lambda _{\Omega }$ are
real. This can be achieved by multiplying the first equation by its
Hermitian conjugate an inserting the second equation ( and viceversa) to get
\begin{eqnarray}
\lambda _{_{\Omega }}H_{\Omega }^{2} &=&\left\vert \lambda _{\Gamma
}\right\vert ^{2}H_{\Gamma }  \label{A3} \\
\lambda _{\Gamma }H_{\Gamma }^{2} &=&\left\vert \lambda _{_{\Omega
}}\right\vert ^{2}H_{\Omega }  \label{A4}
\end{eqnarray}
it follows from these equations that both $\lambda _{_{\Omega }}$ and $\lambda _{\Gamma
}$ should be real since one has two identities among Hermitian matrices.
Now multiplying each of the  Eqs.~(\ref{A1})  on the right 
by $\Gamma _{1}^{\dagger }$ and 
$\Omega _{1}^{\dagger }$ respectively we get 
\begin{equation}
\begin{array}{ccc}
H_{\Omega }H_{\Gamma }=\lambda _{\Gamma }H_{\Gamma } & ; & H_{\Gamma
}H_{\Omega }=\lambda _{\Omega }H_{\Omega }
\end{array}
\label{A5}
\end{equation}
and taking Hermitian conjugates
\begin{equation}
\begin{array}{ccc}
H_{\Gamma }H_{\Omega }=\lambda _{\Gamma }H_{\Gamma } & ; & H_{\Omega
}H_{\Gamma }=\lambda _{\Omega }H_{\Omega }
\end{array}
\label{A5H}
\end{equation}
therefore
\begin{equation}
\lambda _{\Gamma }H_{\Gamma }=\lambda _{\Omega }H_{\Omega }  \label{A6}
\end{equation}
and we conclude that 
\begin{equation}
\left[ H_{\Gamma },H_{\Omega }\right] =0  \label{A7}
\end{equation}
implying that $V_{CKM}=I$ up to permutations of rows or columns. Denoting 
the usual bi-unitary diagonalisation procedure by
\begin{equation}
\begin{array}{ccc}
\Gamma _{1}=V_{L}^{\Gamma }D_{\Gamma }V_{R}^{\Gamma \dag } & ; & \Omega
_{1}=V_{L}^{\Omega }D_{\Omega }V_{R}^{\Omega \dag }
\end{array}
\label{A8}
\end{equation}
from Eq.~(\ref{A7}) we conclude that we can always choose the unitary matrices $
V_{L}^{\Gamma }$ and $V_{L}^{\Omega }$ equal to each other
\begin{equation}
V_{L}^{\Gamma }=V_{L}^{\Omega }  \label{A9}
\end{equation}
and the alignment conditions can be easily reduced to conditions among the
diagonal matrices $D_{\Gamma }$ and $D_{\Omega }$.  From Eq.~(\ref{A5}),
it then follows that:
\begin{equation}
\begin{array}{ccc}
D_{\Omega }^{2}D_{\Gamma }=\lambda _{\Gamma }D_{\Gamma } & ; & D_{\Gamma
}^{2}D_{\Omega }=\lambda _{\Omega }D_{\Omega }
\end{array}
\label{A10}
\end{equation}
It can be checked that there are only two types of solutions. Those with $\lambda
_{\Gamma }$ and $\lambda _{\Omega }$ different from zero (solutions 1,2 and
3) and the remaining ones (solutions 4 and 5 )

\begin{enumerate}
\item $D_{\Gamma }=aP_{3}$ and $D_{\Omega }=\alpha P_{3}$ and changes of $%
P_{3}$ by $P_{2}$ or $P_{1}$

\item $D_{\Gamma }=a\left( I-P_{1}\right) $ and $D_{\Omega }=\alpha \left(
I-P_{1}\right) $ and changes of $P_{1}$ by $P_{2}$ or $P_{3}$

\item $D_{\Gamma }=aI$ and $D_{\Omega }=\alpha I$

\item $D_{\Gamma }=0$ and $D_{\Omega }$ arbitrary and viceversa.

\item $D_{\Gamma }=aP_{i}$ and $D_{\Omega }=\alpha \left( I-P_{i}\right) $
\end{enumerate}

where $P_{i}$ stand for the projection operators
\begin{equation}
\left( P_{i}\right) _{jk}=\delta _{ij}\delta _{ik}  \label{A11}
\end{equation}%
Solutions 2,3 and 4 cannot be good approximations to the actual quark
spectra due to the implied degeneracy. Solution 5 gives rise to $V_{CKM}$ matrix very
different from the identity matrix.  Only solution 1 provides, in leading approximation
the correct pattern of quark masses and mixing. In a suitable weak-basis, this solution
can be written as a democratic matrix $\Delta$.

\section*{Acknowledgments}

This work is partially supported by Spanish MINECO under grant FPA2011-23596,
by Generalitat Valenciana under grant GVPROMETEOII 2014-049
and by Funda\c{c}\~ao para a Ci\^encia e a
Tecnologia (FCT, Portugal) through the projects CERN/FP/123580/2011,
PTDC/FIS-NUC/0548/2012, EXPL/FIS-NUC/0460/2013, and CFTP-FCT Unit 777
(PEst-OE/FIS/UI0777/2013) which are partially funded through POCTI (FEDER),
COMPETE, QREN and EU.

\end{document}